\documentclass[10pt,notitlepage,final]{iopart}

\usepackage{iopams}  
\usepackage{hyperref}
\usepackage{graphicx}
\usepackage[table]{xcolor}
\usepackage{array}
\usepackage{longtable}
\usepackage[english]{babel}

\renewcommand\footnotemark{}

\makeatletter
\def\blfootnote{\xdef\@thefnmark{}\@footnotetext}
\makeatother

\newcommand{\be}{\begin{equation}}
\newcommand{\ee}{  \end{equation}}
\newcommand{\ba}{\begin{eqnarray}}
\newcommand{\ea}{  \end{eqnarray}}

\begin{document}

\title{Emergent properties of nuclei from {\it ab initio}
  coupled-cluster calculations}
\author{G.~Hagen$^{1,2} $, M.~Hjorth-Jensen$^{4,5}$, G.~R.~Jansen$^{1,3}$, and T.~Papenbrock$^{1,2}$}

\address{$^1$Physics Division, Oak Ridge National Laboratory, Oak
  Ridge, TN 37831 USA}

\address{$^2$Department of Physics and Astronomy, University of
  Tennessee, Knoxville, TN 37996, USA}

\address{$^3$National Center for Computational Sciences, Oak Ridge National Laboratory, Oak
  Ridge, TN 37831 USA}

\address{$^4$National Superconducting Cyclotron Laboratory and Department of Physics and Astronomy, Michigan State University, East
  Lansing, MI 48824, USA} 

\address{$^5$Department of Physics,
  University of Oslo, N-0316 Oslo, Norway}

\eads{\mailto{hageng@ornl.gov}}

\begin{abstract}
  Emergent properties such as nuclear saturation and deformation, and
  the effects on shell structure due to the proximity of the scattering
  continuum and particle decay channels are fascinating phenomena in
  atomic nuclei. In recent years, {\it ab initio} approaches to nuclei
  have taken the first steps towards tackling the computational
  challenge of describing these phenomena from Hamiltonians with
  microscopic degrees of freedom. This endeavor is now possible due to
  ideas from effective field theories, novel optimization strategies
  for nuclear interactions, {\it ab initio} methods exhibiting a soft
  scaling with mass number, and ever-increasing computational
  power. This paper reviews some of the recent accomplishments. We
  also present new results. The recently optimized chiral interaction
  NNLO$_{\rm sat}$ is shown to provide an accurate description of both
  charge radii and binding energies in selected light- and medium-mass
  nuclei up to $^{56}$Ni. We derive an efficient scheme for including
  continuum effects in coupled-cluster computations of nuclei based on
  chiral nucleon-nucleon and three-nucleon forces, and present new
  results for unbound states in the neutron-rich isotopes of oxygen
  and calcium. The coupling to the continuum impacts the energies of
  the $J^\pi = {1/2}^-,{3/2}^-,{7/2}^-,{3/2}^+$ states in
  $^{17,23,25}$O, and -- contrary to naive shell-model expectations --
  the level ordering of the $J^\pi = {3/2}^+,{5/2}^+,{9/2}^+$ states
  in $^{53,55,61}$Ca.

\end{abstract}

\submitted{\PS}

\maketitle


Atomic nuclei exhibit a variety of fascinating phenomena that -- on a
microscopic level -- range from single-nucleon motion to the
collective behavior of practically all
nucleons~\cite{bohr1975nobel,mottelson1975nobel}. Many of these can be
viewed as emergent phenomena, that is,  they are very difficult to derive
from (or to understand from) the underlying microscopic Hamiltonian since 
 they differ in scale and complexity from the fundamental
constituents. Here we broadly understand emergent properties as
proposed  by Anderson~\cite{anderson1972}: ``The ability to reduce
everything to simple fundamental laws does not imply the ability to
start from those laws and reconstruct the universe...at each level of
complexity entirely new properties appear, and the understanding of
the new behaviors requires research which I think is as fundamental in
its nature as any other.''

In the physics of atomic nuclei, we mention the following examples of
emergent phenomena: The long-range part of the nuclear interaction
results from pion exchange and is thus a consequence of the
spontaneous breaking of chiral symmetry in quantum chromodynamics
(QCD). Rotations are the lowest excited states in rare earth nuclei
and actinides, as a consequence of the emergent breaking of rotational
symmetry~\cite{bohr1975}. Pairing and nuclear
superfluidity~\cite{bohr1958} results from the emergent breaking of a
unitary symmetry. In neutron-rich nuclei with ground states close to
the breakup thresholds, the interplay between pairing effects and the
coupling to the particle decay channels may cause Borromean properties
and extended halo-like density distributions to emerge
\cite{tanihata1985,zhukov1993, tanihata2013}. Furthermore, the finely
tuned saturation of atomic nuclei as self-bound finite quantum systems
with an approximately constant density can also be viewed as an
emergent phenomenon. Indeed, an inspection of the nucleon-nucleon
interaction would not suggest that nuclear matter saturates, but
repulsive three-nucleon forces (3NFs) combined with attractive
nucleon-nucleon interactions achieve the feat
\cite{bogner2005,hebeler2011,ekstrom2015}. In this article, we also
consider phenomena that emerge in neutron-rich nuclei because these are
inherently open quantum many-body systems with a strong coupling to
the particle continuum and decay channels. Weakly bound and unbound
states are most fascinating properties in neutron-rich nuclei, with
surprising effects on binding and shell structure.

These examples show that much of the rich structure of atomic nuclei
is due to emergent phenomena. Not surprisingly, the microscopic
understanding of emergent phenomena poses significant theoretical
challenges. We note that emergent phenomena are usually collective in
nature and complex at the microscopic level. As a consequence, the
corresponding emergent properties (such as the values for the pion
mass, nuclear moments of inertia, nuclear pairing gaps, the nuclear
saturation energy and density, the properties of halo states for the
examples given in the previous paragraph) are usually finely tuned and
cannot be guessed from the ``natural'' sizes of the parameters of the
underlying microscopic Hamiltonians.

Let us briefly comment on the microscopic approaches to emergent
phenomena in nuclear physics.  Lattice QCD calculations are now
reaching physical pion masses for certain
observables~\cite{green2012,arthur2013,borsanyi2014}. However, there
is disagreement whether nuclear binding
increases~\cite{beane2013,barnea2015} or decreases~\cite{inoue2015}
with increasing pion mass.  First {\it ab initio} computations of
rotational bands in $p$-shell and $sd$-shell nuclei have become
available very
recently~\cite{caprio2013,maris2015,caprio2015,jansen2015,hergert2015}. Here,
the reproduction of moments of inertia for certain deformed $sd$-shell
nuclei is very encouraging, while the accurate computation of
transition strengths from first principles is still an open
problem. {\it Ab initio} computations of superfluid properties have
also been reported recently for nuclei in the oxygen and calcium
region~\cite{soma2013b,signoracci2015}. Here, the accurate
reproduction of pairing gaps and the restoration of particle number
are challenges. Regarding nuclear saturation, only very few nuclear
interactions yield accurate binding energies and radii for
light~\cite{carlson2015} and medium-mass nuclei~\cite{ekstrom2015}
that are also consistent with the empirical saturation point in
nuclear matter. In nuclei heavier than oxygen or so, there seems to be
a tremendous sensitivity to details of the nuclear interaction that
are not probed in few-nucleon systems. Finally, the inclusion of
continuum effects is essential for the level ordering of weakly bound
and unbound states \cite{hagen2012b,langhammer2015}, and the neutron
dripline in exotic nuclei.

Let us also attempt to give a meaningful definition of accuracy in
{\it ab initio} computations. Green's function Monte Carlo methods
computed energy levels of light nuclei to within 2\% accuracy when
compared with data~\cite{pieper2001}. Reference~\cite{hagen2015} lists
a number of heavier nuclei for which binding energies fall within 5\%
of data, and a similar accuracy is possible for radii. In this paper,
we refer to such calculations as being accurate.

The microscopic approach to emergent nuclear pheneomena has become
possible due to advances in computational nuclear physics. 
Chiral effective field theory (EFT), for instance, provides us with
interactions that are consistent with the physics of low-energy QCD,
and the pion mass and pion-nucleon interaction is taken from
data~\cite{ordonez1992,vankolck1994,epelbaum2009,machleidt2011}. Collective
models~\cite{bohr1975,iachello1987} and EFTs of heavy atomic
nuclei~\cite{papenbrock2011,papenbrock2014,coelloperez2015b} are based
on bosonic degrees of freedom, with moments of inertia (or vibrational
frequencies) as input parameters and low-energy constants,
respectively.  Finally, saturation is built into the nuclear shell
model by tuning the frequency of the oscillator basis
accordingly~\cite{brown1988,caurier2005}.

The purpose of this article is to review and discuss recent and new
results on emergent phenomena that now can be approached in {\it ab
  initio} coupled-cluster computations of atomic nuclei. The article
is organized as follows. First, we give a brief theoretical
background. Second, we discuss how nuclear saturation can be viewed as
an emergent property and present the current status of accurate
calculations of nuclear binding energies and radii from chiral
interactions. Third, we discuss how simple and new patterns of
shell-structure emerge in neutron-rich oxygen and calcium isotopes
from complex many-body computations in which effects of both coupling
to the particle continuum and the inclusion of 3NFs are
significant. Fourth, we discuss how to construct non-perturbative
shell-model interactions from coupled-cluster theory starting from
nucleon-nucleon and 3NFs. Using standard shell-model tools we can
describe the emergence of rotational bands from first
principles. Finally, we discuss future perspectives and some of the
challenges that we might expect when extending the nuclear
coupled-cluster program beyond its current reach.

\section{Theoretical background} \label{sec:theory} The
coupled-cluster method was introduced in nuclear physics in the late
1950s by Coester and
K{\"u}mmel~\cite{coester1958,coester1960}. Shortly thereafter it
sparked in quantum chemistry when {\v {C}}{\'\i}{\v z}ek introduced it
in the 1960's as a tool to study  correlated many-electron systems
\cite{cizek1966,cizek1969}. In quantum chemistry it then quickly
became the method of choice for solving many-electron problems. In
nuclear physics on the other hand, it saw only sporadic applications
in the first few decades after its
inception~\cite{kuemmel1978,bishop1978,bishop1991}. During the last
decade or so, coupled-cluster theory has been revitalized in nuclear
physics~\cite{mihaila2000b,dean2004,wloch2005,hagen2007b,hagen2010b,roth2011a}
and has been used to compute nuclei from scratch based on modern
forces from chiral EFT, see Ref.~\cite{hagen2014} for a recent
review. Its role has also changed from being mostly post-dictive to
experiment to also having predictive capabilities. Examples are the
prediction of the energy of the first excited $2^+$ state in the
neutron-rich isotope $^{54}$Ca~\cite{hagen2012b}, experimentally
confirmed by Steppenbeck {\it et
al.}~\cite{steppenbeck2013,steppenbeck2013b}, and the predictions of
the dipole polarizability and the neutron distribution of
$^{48}$Ca \cite{hagen2015}.

Let us briefly discuss some technical details of the coupled-cluster
method and refer the reader to Ref.~\cite{bartlett2007} for a
comprehensive review.  Coupled-cluster theory can be viewed as a
(non-unitary) similarity transformation of the Hamiltonian such that
the Hartree-Fock reference becomes an exact eigenstate in a Hilbert
space limited to $n$-particle--$n$-hole ($n$p-$n$h) excitations of the
reference state. The similarity transformation is generated by the
exponentiation of particle-hole excitation operators up to the
$n$p-$n$h level. The workhorse is the coupled-cluster singles doubles
(CCSD) approximation corresponding to $n=2$, and triples $n=3$
excitations are included perturbatively.  The method has a polynomial
cost in mass number and size of the single-particle basis, and
Hartree-Fock references corresponding to closed (sub-)shell nuclei are
computationally least expensive.  The similarity-transformed
Hamiltonian is at the center of the coupled-cluster method. Its
diagonalization (via so-called ``equation-of-motion methods'') in a
Fock-space basis consisting of generalized particle-hole excitations
yields excited states in the reference nucleus and states in
neighboring nuclei that differ by one or two units in particle numbers
or in the difference between neutron and proton numbers.

We perform coupled-cluster calculations based on the intrinsic
Hamiltonian, 
\begin{equation}
  \label{intham}
  \hat{H} = \sum_{i<j}\left({({\bf p}_i-{\bf p}_j)^2\over 2mA} + \hat{V}
    _{NN}^{(i,j)}\right) + \sum_{ i<j<k}\hat{V}_{\rm 3N}^{(i,j,k)},
\end{equation}
where $\hat{V}_{NN}$ is the nucleon-nucleon and $\hat{V}_{\rm 3N}$ is
the three-nucleon interaction. In coupled-cluster computations of
nuclei the Hamiltonian enters in normal-ordered form, i.e.
\begin{eqnarray}
\nonumber
H_{N} & = & \sum_{pq}\langle p \vert f \vert q \rangle :a^\dagger_{p}
a_q: \nonumber \\ & + &
\frac{1}{4}\sum_{pqrs}\langle pq \vert v \vert rs \rangle
:a^\dagger_{p} a^\dagger_q a_{s}a_r: \\ 
& + &  
\nonumber 
\frac{1}{36}\sum_{pqrstu}\langle pqr \vert w \vert stu \rangle
:a^\dagger_{p} a^\dagger_q a^\dagger_{r}a_ua_ta_s:, 
\label{eq:ham_N}
\end{eqnarray}
with the normal-ordered string of creation and annihilation operators
$:a^\dagger_{p} \ldots a_{p'}\ldots: $. The normal-ordered one-body
Fock matrix is
\begin{equation}
  \label{eq:fock} 
  \langle p \vert f \vert q \rangle = \langle p \vert
  t \vert q \rangle + \sum_{i}\langle p i \vert V_{\mathrm NN} \vert q
  i \rangle + \frac{1}{2}\sum_{ij} \langle pij \vert
  V_{\mathrm{3NF}} \vert qij \rangle,
\end{equation}
the normal-ordered two-body operator is given by 
\begin{equation}\label{eq:no2}
  \langle pq \vert v \vert rs \rangle  =   
  \langle pq \vert V_{\mathrm NN} \vert rs \rangle
   +  
  \sum_i \langle pqi \vert V_{\mathrm{3NF}} \vert irs \rangle,
\end{equation}
and the normal-ordered three-body operator $w$ is given by 
\begin{equation}
\label{no3}
  \langle pqr \vert w \vert stu \rangle = 
   \langle pqr \vert V_{\mathrm{3NF}} \vert stu \rangle. 
\end{equation}
Coupled-cluster calculations that include 3NFs are computationally
very expensive \cite{hagen2007a,binder2013}, and we therefore use the
normal-ordered two-body approximation of the 3NF. This amounts to
neglecting the residual three-body term given in Eq.~(\ref{no3}). This
approximation is accurate in light and medium mass
nuclei~\cite{hagen2007a,roth2012,binder2013}.

We note that the coupled-cluster method preserves the translational
invariance of the Hamiltonian to a large degree of accuracy, because
the coupled-cluster wave function factors into an intrinsic wave
function and a Gaussian center-of-mass wave
function~\cite{hagen2009a,hagen2010b,jansen2012}. This property is not
limited to the coupled-cluster method but is also demonstrated by
other methods such as the in-medium similarity renormalization group
(IMSRG) approach, see for example Refs.~\cite{morris2015,hergert2015}.

\section{Nuclear saturation} 
Atomic nuclei along the valley of $\beta$ stability are bound by about
8~MeV per nucleon, and charge radii scale approximately as $R_c\approx
1.2 A^{1/3}$~fm, with $A$ denoting the mass number.  The accurate
computation of nuclear binding energies and radii from microscopic
Hamiltonians has been a long-standing challenge in nuclear theory.
For the light ($p$-shell) nuclei this challenge was successfully
addressed by quantum Monte Carlo computations~\cite{carlson2015}. For
interactions based on chiral EFT, radii and binding energies are well
reproduced for light nuclei, while oxygen isotopes and heavier nuclei
typically exhibit too much binding and too small radii, see
Refs.~\cite{binder2013b,lahde2014} for examples, and
Ref.~\cite{ekstrom2015} for a compilation of results.

We note that nuclear saturation is a finely tuned property, and give
several examples. Hebeler {\it et al.}~\cite{hebeler2011} considered
families of similarity-renormalization group (SRG) transformed $NN$
interactions \cite{bogner2007} from chiral EFT with ``bare'' 3NFs
adjusted to the radii and binding energies of $^3$H and $^4$He. While
the different interactions agree on the binding energy of the $^4$He
nucleus within 1\%, the corresponding Fermi momenta and binding
energies at the saturation point of nuclear matter vary by about 10\%
and 25\%, respectively. Likewise, Carlsson {\it et
  al.}~\cite{carlsson2015} found that different parametrizations of
chiral EFT obtained from varying the chiral cutoff and the maximum
nucleon-nucleon scattering energy used in the optimization, gives a
spread in the binding energy of $^4$He of about 7\%, while the
corresponding spread in $^{16}$O is about 40\%.
In nuclear lattice EFT calculations based on chiral nucleon-nucleon
and 3NFs, a repulsive phenomenological four-body contact was added to
the Hamiltonian to compensate for the overbinding observed in nuclei
heavier than $^{16}$O \cite{lahde2014}. Very recently, lattice EFT
computations proposed that nuclear saturation is very sensitive to the
non-locality of the nucleon-nucleon interaction already at leading
order, and that nuclei might be close to a quantum phase
transition~\cite{elhatisari2016}. To obtain proper saturation
properties, $\alpha$-$\alpha$ scattering (an eight-body phenomenon)
was used to tune the lattice EFT interaction.
 
We recall that the family of interactions from Ref.~\cite{hebeler2011}
differ in their high-momentum cutoff. This suggests that the
saturation point in nuclear matter is sensitive to details of the
chiral interaction.  Consistent with this, coupled-cluster
calculations of nuclear matter with interactions from chiral EFT
showed that the saturation point is very sensitive to whether the
employed high-momentum regulator functions are local or
non-local~\cite{hagen2013b}. On the one hand, these findings seem to
be unexpected from the view of an EFT. After all, differences in
high-momentum aspects are supposed to be higher-order effects in the
chiral power counting. On the other hand, we remind the reader that
the power counting is in the potential energy, and not in the total
binding energy. Thus, it might be that variation in the binding energy
is consistent with (3NF) interactions at N$^2$LO. Finally, we note
that neutron matter is less sensitive to details of the high-momentum
regulators~\cite{hebeler2013,hebeler2015}, presumably because only the
two-pion exchange 3NF enters (for non-local regulators). For
interactions from chiral EFT, the saturation mechanism consists of
relatively soft $NN$ forces that -- taken alone -- yield too much
binding and too small radii. The 3NFs act repulsively in matter and
heavier nuclei, playing  thereby an important role in  nuclear 
saturation~\cite{bogner2005,hebeler2011,ekstrom2015}.

Reference~\cite{hagen2013b} also demonstrated that no adjustment of
the short-range parts of the leading 3NFs is able to simultaneously
reproduce the binding energy of $^3$H and the saturation point of
nuclear matter (within a 5\% uncertainty). A similar result has been
found for the Argonne V18 (AV18) $NN$ interaction~\cite{wiringa1995}
and a family of 3NF models~\cite{logoteta2015}. In
Refs.~\cite{maris2013,carlson2015} it was also noted that AV18 with
the Illinois 3NFs produce accurate results for binding energies and
spectra of $p$-shell nuclei while less so for neutron matter. On the
other hand, AV18 with the Urbana 3NFs produce good results for neutron
matter while less accurate results for light $p$-shell nuclei.

\begin{figure}[htb]
\includegraphics[width=0.85\textwidth]{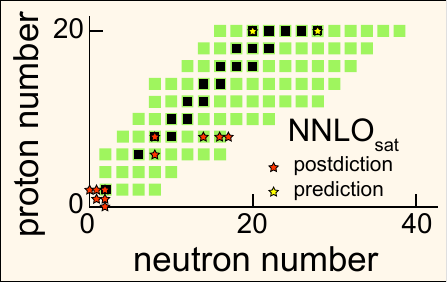}
\caption{(Color online) Segr{\`e} chart of light even-even nuclei
  (green and black squares). $\beta$-stable nuclei are denoted as black
  squares. Red stars mark nuclei included in the optimization of the
  chiral interaction NNLO$_{\rm sat}$. Yellow stars mark nuclei for
  which predictions were made in Ref.~\cite{hagen2015}.}
  \label{NNLOsat_strategy}
\end{figure}

In Ref.~\cite{ekstrom2015} this challenge was addressed by including
data from ground-state energies and charge radii of $^2$H, $^3$H,
$^3$He, $^4$He, $^{14}$C, $^{16}$O and ground-state energies of
$^{22,24,25}$O into the optimization of the nuclear interaction
NNLO$_{\rm sat}$ from chiral EFT at
NNLO. Figure~\ref{NNLOsat_strategy} shows the nuclei employed in the
optimization as red stars on the Segr{\`e} chart. 

As a novelty, the $NN$ and 3NF contributions of the chiral force
NNLO$_{\rm sat}$ were also optimized simultaneously (and not
sequentially). This approach is consistent with the inclusion of light
nuclei into the optimization and with an adjustment of low-energy constants (LECs)
order-by-order in the power counting.  We refer the reader to
Ref.~\cite{carlsson2015} for a detailed comparison and discussion of
sequentially and simultaneously optimized chiral interactions.

The inclusion of light nuclei beyond the three-nucleon bound states in
the optimization of the nuclear interactions breaks with the
traditional approach. There are additional arguments in support of
this procedure. We note that the isospin $T=3/2$ component of the 3NF
is not constrained by bound states in three- or four-nucleon
systems. We also note that EFTs are concerned only with the
description of low-energy data and observables~\cite{lepage1997}. It
is therefore not necessary to limit the determination of LECs to the
simplest systems imaginable. In EFTs of magnets, for instance, the
relevant LECs are not determined in the two-body system but rather
from bulk properties at low
energies~\cite{leutwyler1994,roman1999,kampfer2005}. Finally, we note
that the most accurate computations of light nuclei up to mass number
$A\le 12$~\cite{carlson2015} employ 3NFs that have been adjusted to 20
states in nuclei with mass numbers $A\le 10$
\cite{pieper2001b,carlson2015}.

Though they result from a non-perturbative solution of a quantum
many-body problem, ground-state energies and charge radii of finite
nuclei are truly low-energy observables -- and their inclusion in the
optimization of LECs is thus consistent with EFT
ideas. Reference~\cite{binder2015} shows that the binding energies of
$^3$H, $^4$He, and $^6$Li exhibit the same convergence pattern with
increasing order of the chiral power counting as scattering
observables at low energies of 10~MeV.

\begin{figure}[htb]
\includegraphics[width=0.85\textwidth]{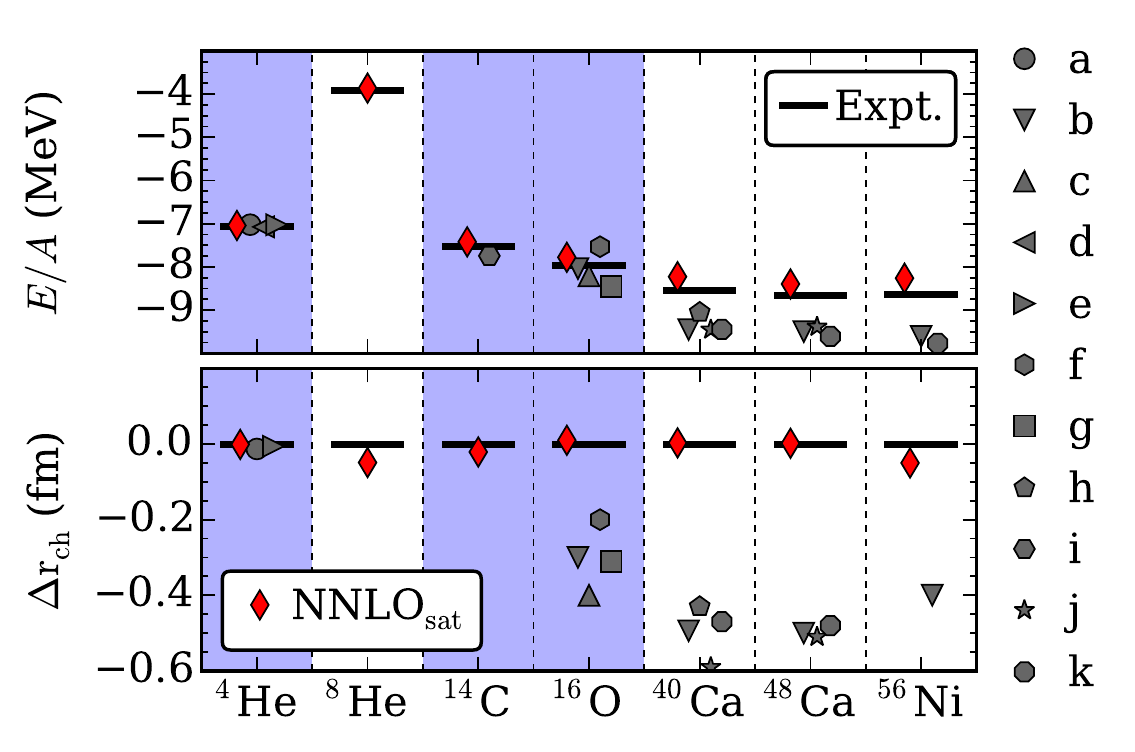}
\caption{(Color online) Ground-state energies per nucleon (top) and
  differences between theoretical and experimental charge radii
  (bottom) for selected nuclei and results from {\it ab initio}
  computations. The red diamonds mark results based on the chiral
  interaction NNLO$_{\rm sat}$. The blue shaded columns indicate which
  nuclei where included in the optimization of NNLO$_{\rm sat}$, while
  the white columns are predictions. References:
  $a$~\cite{navratil2007b,jurgenson2011}, $b$~\cite{binder2013b},
  $c$~\cite{epelbaum2014}, $d$~\cite{epelbaum2012},
  $e$~\cite{maris2014}, $f$~\cite{wloch2005}, $g$~\cite{hagen2012a},
  $h$~\cite{bacca2014},
  $i$~\cite{maris2011},$j$~\cite{soma2013b},$k$~\cite{hergert2014}.}
  \label{NNLOsat2}
\end{figure}

Figure~\ref{NNLOsat2} shows the results for ground-state energies per
particle and differences between theoretical and experimental charge
radii of selected light and medium-mass nuclei computed with
NNLO$_{\rm sat}$ and compares them to other {\it ab initio}
computations. Here the results up to $^{40}$Ca are from
Ref.~\cite{ekstrom2015}, the results for $^{48}$Ca from
Ref.~\cite{hagen2015}, while the results for $^{56}$Ni are new. For
the last two nuclei, the ground-state energy is computed using the
CCSD with perturbative $\Lambda$-triples ($\Lambda$-CCSD(T))
\cite{taube2008}, while for the lighter nuclei we also added a
perturbative estimate for the residual 3NF term in Eq.~(\ref{no3})
(yielding about 1\% in additional binding energy). The computed
binding energy per particle for $^{56}$Ni is 8.26~MeV slightly less
bound than the experimental value of 8.64~MeV. The charge radius of
$^{56}$Ni is not measured. Therefore we compare the computed charge
radius of 3.72~fm to the known charge radius of 3.77~fm for $^{58}$Ni
(which is probably larger than what is expected for $^{56}$Ni.). The
coupled-cluster calculations of $^{56}$Ni were performed in the same
model space as Ref.~\cite{hagen2015}. We see that NNLO$_{\rm sat}$
provides us with an improved description of the saturation properties
in medium-mass nuclei.

Recently, Hagen~{\it et al.}~\cite{hagen2015} employed NNLO$_{\rm
  sat}$ for the computation of the neutron density in $^{48}$Ca. The
good agreement between theoretical results and data for the charge
radii of $^{40,48}$Ca validated this approach, see
Fig.~\ref{NNLOsat2}. The radius $R_n$ of the point-neutron density was
predicted as $3.47 \lesssim R_n \lesssim 3.60$~fm. Here, the
theoretical uncertainties were estimated by correlating the
theoretical results from NNLO$_{\rm sat}$ and a family of chiral
interactions~\cite{hebeler2011} with the precisely known experimental
charge radius. The strong correlation between the charge radii and
point-neutron radii distinguishes {\it ab initio} results from nuclear
density functional theory (DFT). As a consequence, {\it ab initio}
results for the neutron skin, i.e. the difference between
point-neutron and point-proton radii, exhibit almost no dependence of
the employed chiral interaction. This is in contrast to results from
nuclear DFT; the latter also exceed the former significantly.

As we have seen, chiral interactions with acceptable saturation
properties have become available in recent
years~\cite{hebeler2011,ekstrom2015}, and they allow us to address
burning questions in the physics of atomic nuclei. The extension of
this program to even heavier nuclei is underway~\cite{binder2015b}.

\section{Continuum effects}
Recently, efforts have been made in formulating an \emph{ab initio}
theory that unifies both structure and reactions in nuclei
\cite{barbieri2005,hagen2007d,quaglioni2008,hagen2012c,elhatisari2015,navratil2016}. To
describe reactions in nuclei one needs to account for decay channels
in which coupling to both bound- and scattering states are
important. The coupling to the continuum impacts also the level
ordering of unbound states and shell-structure in nuclei.
Understanding how shell structure evolve in neutron-rich nuclei is of
great experimental and theoretical interest, in particular since shell
structures impact the limits of stability known as the dripline (that
is the limit where adding another neutron or proton to a bound isotope
does not produce a particle stable ground-state), and thereby the
number of nuclei that can exist. As the neutron-to-proton ratio varies
it has been found in several isotopes that the magic numbers
$2,8,20,28,50,\ldots$ of the shell-model of Goeppert-Mayer and Jensen
can be less magic than expected, while the appearance of new magic
numbers in other isotopes have been observed. In this section we
discuss and present new results for unbound states in the neutron-rich
oxygen and calcium isotopes, with an emphasis on the role of coupling
to the particle continuum and state-of-the-art chiral nucleon-nucleon
and three-nucleon interactions.

\subsection{Coupled-cluster calculations based on a Gamow-Hartree-Fock
  basis}
As discussed in the previous section, coupled-cluster calculations
based on chiral nucleon-nucleon and 3NFs can now accurately describe
bulk properties such as binding energies and radii of light- and
medium-mass nuclei. To provide experiment with reliable predictions
for exotic nuclei, it is also necessary that the coupling to the
scattering continuum is properly taken into account. Here we briefly
outline and discuss how the coupling to the particle-continuum can be
taken into account in \emph{ab initio} coupled-cluster calculations of
nuclei based on chiral nucleon-nucleon and 3NFs.

\begin{figure*}[tbh]
\includegraphics[width=1.0\textwidth]{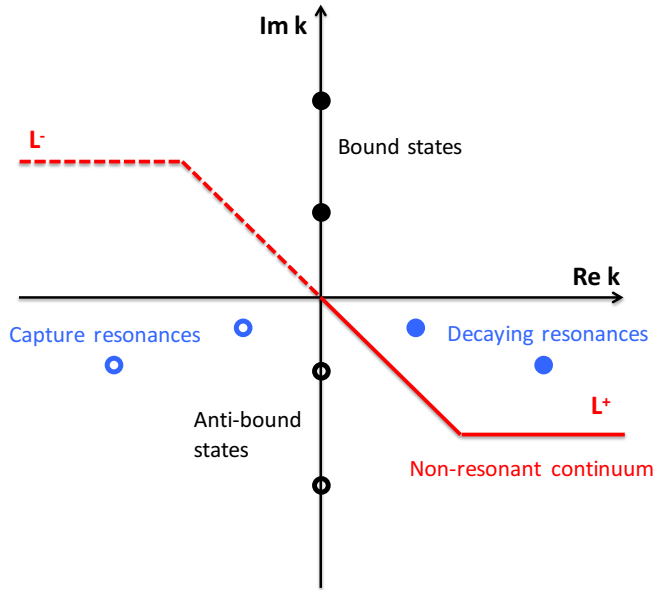}
\caption{(Color online) Poles of the $S$-matrix in the complex
  $k$-plane, and the contour $L^+$ which defines the non-resonant
  continuum (see Ref.~\cite{hagen2004} for details).}
  \label{smat_poles}
\end{figure*}

Similar to what is done in the Gamow-shell-model
\cite{michel2002,idbetan2002,hagen2006b}, we use a Berggren basis
\cite{berggren1968,berggren1971,lind1993} as a starting point for our
continuum coupled-cluster calculations
\cite{hagen2007d,hagen2010a}. As depicted in Fig.~\ref{smat_poles} one
can choose an appropriate contour $L^+$ in the complex $k$-plane such
that the usual completeness relation can be written as a discrete sum
over bound- and resonant states and with an integral over the
non-resonant continuum $L^+$ \cite{hagen2004}. This is the
Berggren completeness relation,
\begin{equation}
\label{eq:unity2}
{\bf 1} = \sum _{n\in \bf{C}}\vert\psi_{nl}\rangle\langle\psi_{nl}^{*}\vert + 
\int_{L^{+}} dk k^2\vert\psi_{l}(k)\rangle\langle\psi_{l}^{*}(k)\vert.
\end{equation} 
In numerical applications the integral over $L^+$ is discretized using
a suitable quadrature rule such as Gauss-Legendre, and converged
results are usually obtained with 30-40 quadrature points and a
maximum real momentum of $\mathrm{Re}[k] = 4-5 \mathrm{fm}^{-1}$.

To obtain the Hamiltonian in the Berggren basis, $\vert a\rangle$, we
utilize a finite expansion over harmonic oscillator states, $\vert
\alpha \rangle $, i.e.
\begin{equation}
  \langle ab\vert V_{NN}\vert cd\rangle \approx
  \sum_{\alpha \beta \gamma \delta}^{n_{\rm max}} \langle ab\vert \alpha \beta \rangle
  \langle \alpha \beta \vert V_{NN}\vert \gamma \delta \rangle \langle \gamma \delta \vert cd \rangle,
  \label{nn_exp}
\end{equation}
which in the limit $n_{\rm max} \rightarrow \infty $ is exact, while
the kinetic energy is evaluated directly in the Berggren basis (see
Ref.~\cite{hagen2007d} for more details). We obtain a
Gamow-Hartree-Fock basis \cite{michel2004,hagen2007d} by solving the
Hartree-Fock equations in a mixed representation of
harmonic-oscillator and Berggren basis states. This treats bound,
scattering, and resonant states on an equal footing.

To construct a Gamow-Hartree-Fock basis starting from the Hamiltonian
Eq.~(\ref{intham}), the 3NF and the nucleon-nucleon interaction need
to be represented in the Berggren basis. Generalizing the expansion
given in Eq.~(\ref{nn_exp}) to include 3NFs is formally
straightforward, but it represents a significant increase in the
computational cost and memory. Below we show how a Gamow-Hartree-Fock
basis can be computed efficiently. Let us start with the one-body
Hartree-Fock Hamiltonian in the Berggren basis,

\begin{equation} 
\langle a \vert h_{\rm HF} \vert c\rangle = 
\langle a \vert t \vert c\rangle + 
\langle a \vert v_{ NN} \vert c\rangle + 
\langle a \vert v_{ 3N} \vert c\rangle,
\label{eq:hf1}
\end{equation}
consisting of a one-body kinetic energy term, and one-body potential
terms arising from folding the nucleon-nucleon interaction and the
3NF with one-body density matrices, that is
\begin{eqnarray} 
\langle a \vert v_{ NN} \vert c\rangle = \sum_J\sum_{pq} \left(
        {2J+1\over 2j_a+1}\right) \langle ap \vert V_{NN} - { {\bf
            p}_1\cdot{\bf p}_2\over 2mA} \vert cq \rangle^{(J)}\langle
        q\vert \rho\vert p\rangle, \\ \langle a \vert v_{3N} \vert
        c\rangle = {1\over 2}\sum_J\sum_{pqrs} \left( {2J+1\over
          2j_a+1}\right) \langle apq \vert V_{3N} \vert c
        rs\rangle^{(J)}\langle r\vert \rho\vert p\rangle\langle s\vert
        \rho\vert q\rangle.
\end{eqnarray}  
Note that $\langle a \vert v_{ NN} \vert c\rangle $ contains
contributions from both the nucleon-nucleon interaction and from the
two-body part of the intrinsic kinetic energy. The one-body density
matrices $\langle p \vert \rho \vert q\rangle $ are constructed from
the eigenvectors obtained by diagonalizing Eq.~(\ref{eq:hf1}). The
computation of $\langle a \vert v_{3N} \vert c\rangle $ is expensive
in the Berggren basis, however we can circumvent the problem of
computing matrix elements of the 3NF directly in the Berggren basis by
first writing,
\begin{eqnarray} 
\nonumber \langle \alpha \vert v_{3N} \vert \gamma \rangle = {1\over
  2}\sum_J\sum_{pqrs}\sum_{\kappa\lambda\mu\nu} \left( {2J+1\over
  2j_a+1}\right) \langle \alpha \kappa\lambda \vert V_{3N} \vert
\gamma \mu\nu \rangle^{(J)} \\ \nonumber \langle p\vert \kappa\rangle
\langle q\vert \gamma\rangle \langle \mu\vert r\rangle \langle
\nu\vert s\rangle \langle r\vert \rho\vert p\rangle\langle s\vert
\rho\vert q\rangle \\ = {1\over 2}\sum_J\sum_{\kappa\lambda\mu\nu}
\left( {2J+1\over 2j_a+1}\right) \langle \alpha \kappa\lambda \vert
V_{3N} \vert \gamma \mu\nu \rangle^{(J)} \langle \mu \vert \rho\vert
\kappa \rangle\langle \nu\vert \rho\vert \lambda \rangle .
\label{eq:hf2}
\end{eqnarray} 
Here, $\langle a \vert \alpha\rangle$ are radial overlap functions
between a Berggren and a harmonic-oscillator basis state, and in the
last term we introduced projections of the density matrices onto a
finite harmonic-oscillator space, i.e.
\begin{equation}
\langle \alpha \vert \rho \vert \beta \rangle = \sum_{ab} \langle
\alpha\vert a \rangle \langle a \vert \rho \vert b \rangle \langle
 b \vert \beta \rangle.
\end{equation}
Finally, in order to obtain the one-body matrix elements 
$\langle a \vert v_{3N} \vert c\rangle $, we introduce the expansion, 
\begin{equation}
\langle a \vert v_{3N} \vert c\rangle = \sum_{\alpha \beta} \langle a
\vert \alpha\rangle \langle \alpha \vert v_{3N} \vert \beta \rangle
\langle \beta \vert b \rangle.
\label{eq:hf3}
\end{equation}
Using this strategy we can compute a Gamow-Hartree-Fock basis from
nucleon-nucleon and 3NFs self-consistently by never having to
represent the 3NF directly in the Berggren basis. The only requirement
is that the one-body density matrices defined in the Berggren basis
are projected onto a finite harmonic oscillator space after each
Hartree-Fock iteration, then proceed by computing the 3NF contribution
in Eq.~(\ref{eq:hf2}), and finally projecting the $v_{3N}$ term onto the
Berggren basis via Eq.~(\ref{eq:hf3}). In practice one verifies that the
results are converged with respect to the size of the finite harmonic
oscillator space used. Once the Gamow-Hartree-Fock basis is
self-consistently determined, the normal-ordered one- and two-body
matrix elements given in Eqs.~(\ref{eq:fock}) and (\ref{eq:no2}) are
evaluated in this basis.

\subsection{Unbound states in neutron-rich
  oxygen isotopes}
The oxygen isotopes have been extensively studied during the  last decade. Here new (sub-) shell
closures at $N=14$ and $N=16$ have been observed experimentally
\cite{stanoiu2004,kanungo2009,hoffman2009}, while the $N=20$ magic
number is weakened and disappears along the isotone chain starting
from $^{40}$Ca towards the island of inversion nuclei $^{32}$Mg and
$^{30}$Ne, where the onset of large deformations is observed
\cite{motobayashi1995}. Furthermore, $^{24}$O is the last bound
isotope with the ground-state of $^{25}$O being particle unstable at
the neutron decay energy $770$~keV and with a width of $172(30)$keV
\cite{hoffman2008}. This has become known as the oxygen anomaly, since
by adding just one more proton pushes the neutron dripline
considerably further out with $^{31}$F being the last known bound
fluorine isotope \cite{sakurai1999}. This oxygen anomaly was given a
theoretical explanation in Ref.~\cite{otsuka2010}, where it was shown
in shell-model calculations that 3NFs act repulsively and correctly
set the dripline at $^{24}$O. Perhaps even more interesting than
$^{25}$O is its neighbor $^{26}$O. This nucleus has recently been
observed at a decay energy of $150^{+50}_{-150}$keV and with a width
of $5$keV. It decays by two-neutron emission
\cite{lunderberg2012}. Few-body models of $^{26}$O have empasized the
role of di-neutron correlations in the two-neutron direct decay of
this nucleus, and based on the experimentally determined width of
$5$keV set an upper bound of $1$keV for the decay energy
\cite{grigorenko2013}, in other words right at the two-neutron
emission threshold. More precise data for this nucleus will challenge and 
impact on our theoretical understanding of two-neutron decays in
neutron-rich nuclei. 

The neutron-rich nucleus $^{28}$O is even more exotic. In the standard
shell-model picture this nucleus is a doubly magic nucleus with $N=20$
and $Z=8$ shell closures. \emph{Ab initio} calculations of this
nucleus based on nucleon-nucleon interactions and 3NFs all predict
that $^{28}$O is particle unstable with respect to $^{24}$O
\cite{otsuka2010,hergert2013,cipollone2013}. However, none of these
calculations take into account the coupling to the particle continuum
which impacts loosely bound and unbound states.

In Ref.~\cite{hagen2012a} the role of coupling to the
particle-continuum was investigated on low-lying excited states in
neutron-rich oxygen isotopes. The effects of 3NFs were included in
terms of a schematic density dependent correction to the chiral
nucleon-nucleon interaction N$^3$LO \cite{entem2003} by
normal-ordering the 3NF in symmetric nuclear matter. It was found that
the coupling to the particle continuum had a modest effect of at most
a few hundred keV's on states that are dominated by $sd$-shell
configurations. The effect of the particle continuum on
negative-parity states and possible intruder states from the
$pf$-shell will be addressed below.

Phenomenological shell-model approaches that properly take into
account the coupling to the particle continuum also exist. The
Gamow-shell-model \cite{michel2002,idbetan2002} has been successfully
applied to unbound states in selected isotopes of helium and oxygen.
In Ref.~\cite{volya2005} the particle continuum was included in a
unified shell-model approach based on the phenomenological USD
interaction \cite{brown1988} with accurate predictions for
$^{25,26}$O, while $^{28}$O was predicted to be unbound with a decay
energy of $1.44$MeV.

\begin{figure*}[t]
\includegraphics[width=1.0\textwidth]{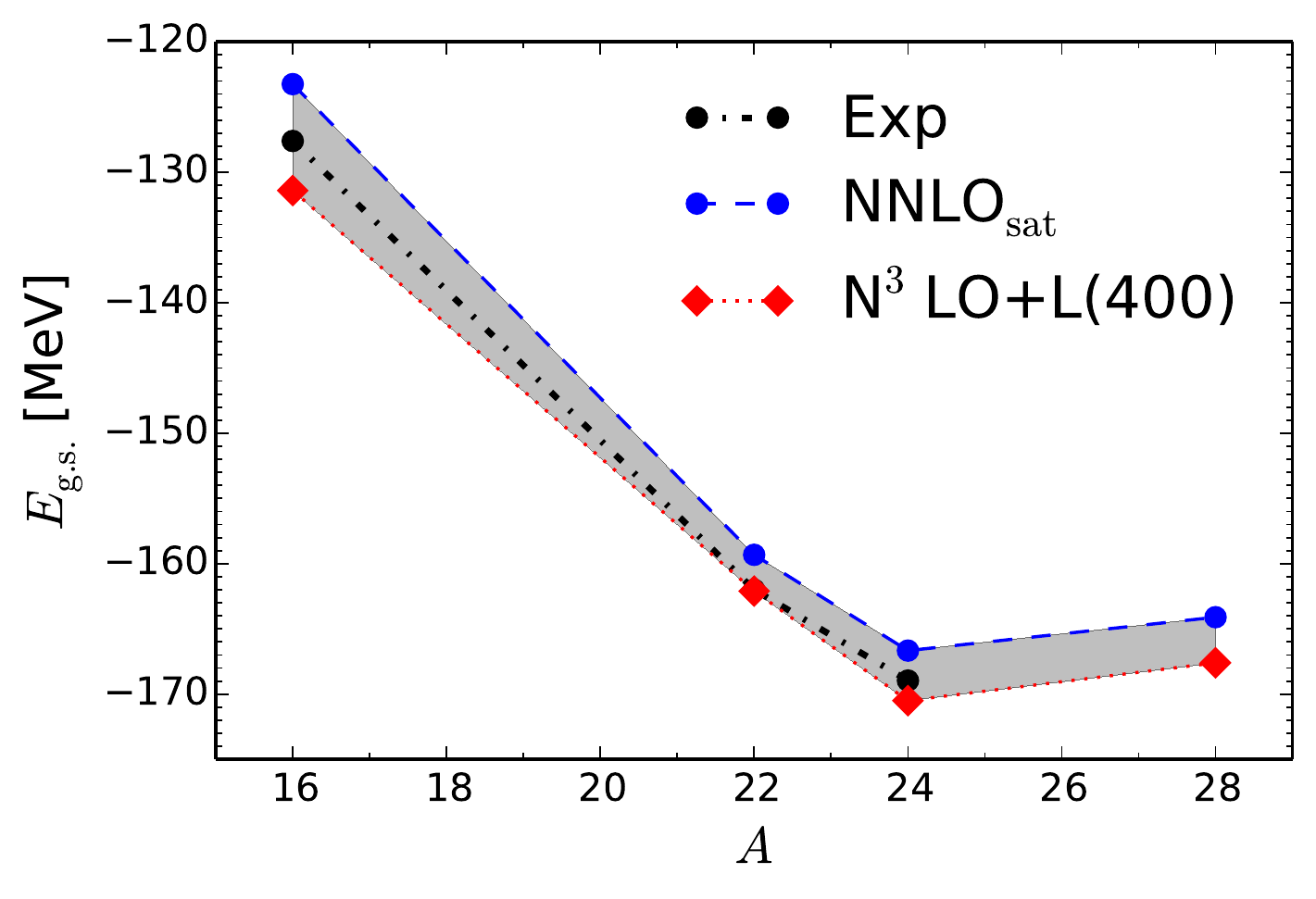}
\caption{(Color online) Ground-state energies of the oxygen isotopes
  $^{16,22,24,28}$O computed from $\Lambda$-CCSD(T) using two
  NNLO$_{\rm sat}$ (blue solid circlies) and N$^3$LO + L(400) (red
  solid diamonds), and compared to experiment (black dashed-dotted
  line). The gray band estimates systematic uncertainties of the
  nuclear interaction.}
  \label{Ox_BE}
\end{figure*}

Figure \ref{Ox_BE} shows the computed binding energies of
$^{16,22,24,28}$O using the $\Lambda$-CCSD(T) approximation from two
different chiral interactions. Experimental data are included in the
figure. First we consider the recently optimized NNLO$_{\rm sat}$
\cite{ekstrom2015}, and then the chiral nucleon-nucleon interaction at
N$^3$LO~\cite{entem2003} with the 3NF at N$^2$LO using a local
regulator with cutoff $\Lambda_{\rm 3N} = 400$ MeV (which we denote as
N$^3$LO + L(400)) \cite{navratil2007,roth2012}. We note that for
NNLO$_{\rm sat}$ the binding energies of $^{16,22,24}$O were included
in the fit, while for N$^3$LO + L(400) only data up to $^4$He were
included in the fit of the low-energy constants. The Hamiltonian
N$^3$LO + L(400) has been shown to produce accurate results for both
binding energies and spectra in and around the oxygen isotope chain
\cite{hergert2013,cipollone2013,bogner2014,jansen2014}. The
model-space used is a Gamow-Hartree-Fock basis built from a mixed
representation of $N_{\rm max} = 2n+l = 14$ harmonic-oscillator basis
functions and a Berggren basis for the $d_{3/2}$ neutron partial
wave. The oscillator frequencies were set to $\hbar\Omega = 22$~MeV
for NNLO$_{\rm sat}$, and $\hbar\Omega = 20$~MeV for N$^3$LO + L(400),
respectively. For the 3NFs we used the additional energy cuts
$E_{3\rm{max}}=N_1+N_2+N_3 \leq 16$ for NNLO$_{\rm sat}$, and
$E_{3\rm{max}}=N_1+N_2+N_3 \leq 14$ for N$^3$LO + L(400),
respectively. In Refs.~\cite{hergert2013,bogner2014} it was found that
this model space yields practically converged results (0.1\% for
binding energies and 100-200~keV for spectra) in oxygen isotopes for
the N$^3$LO + L(400) interaction. For both Hamiltonians $^{28}$O is
predicted to be unbound with respect to $^{24}$O by 2.5 for NNLO$_{\rm
  sat}$ and 2.9~MeV for N$^3$LO + L(400), respectively. We found that
the inclusion of the continuum for the $d_{3/2}$ partial wave adds
about 1~MeV of additonal binding compared to using a harmonic
oscillator basis for this partial wave.

The exotic nucleus $^{28}$O is close to the known island of inversion
region, where intruder states from the $pf$-shell, deformation and
continuum couplings play an important role for the structure of
ground- and low-lying excited states
\cite{zhang2014,xu2015}. Furthermore, relativistic mean-field
calculations of $^{31}$Ne found that the $pf$-orbitals are inverted
due to continuum coupling even in the case of a spherical or slightly
deformed $^{31}$Ne \cite{zhang2014}. It is therefore reasonable to
expect that configuration mixing from the $pf$-shell and continuum
coupling will play a role on the structure of $^{28}$O as well. In
order to cast some light on how important configuration mixing with
the $pf$-shell may be on the structure of neutron-rich oxygen isotopes
we study the evolution of the unbound $J^\pi =
{1/2}^-,{3/2}^-,{7/2}^-$ states and the unbound $J^\pi = {3/2}^+$
state that are dominated by one-particle configurations in the odd
oxygen isotopes $^{17,23,25}$O. These states can be computed in the
particle-attached equation-of-motion coupled-cluster (PA-EOM-CC)
method \cite{gour2006}, and we include continuum effects by using a
Berggren basis for the $d_{3/2},p_{1/2},p_{3/2},f_{7/2}$ partial
waves. The PA-EOM-CC method was previously used to compute resonances
in $A=17$ nuclei using nucleon-nucleon forces only \cite{hagen2010a},
while here we for the first time we include also 3NFs and compute
negative-parity states.

\begin{figure*}[t]
\includegraphics[width=1.0\textwidth]{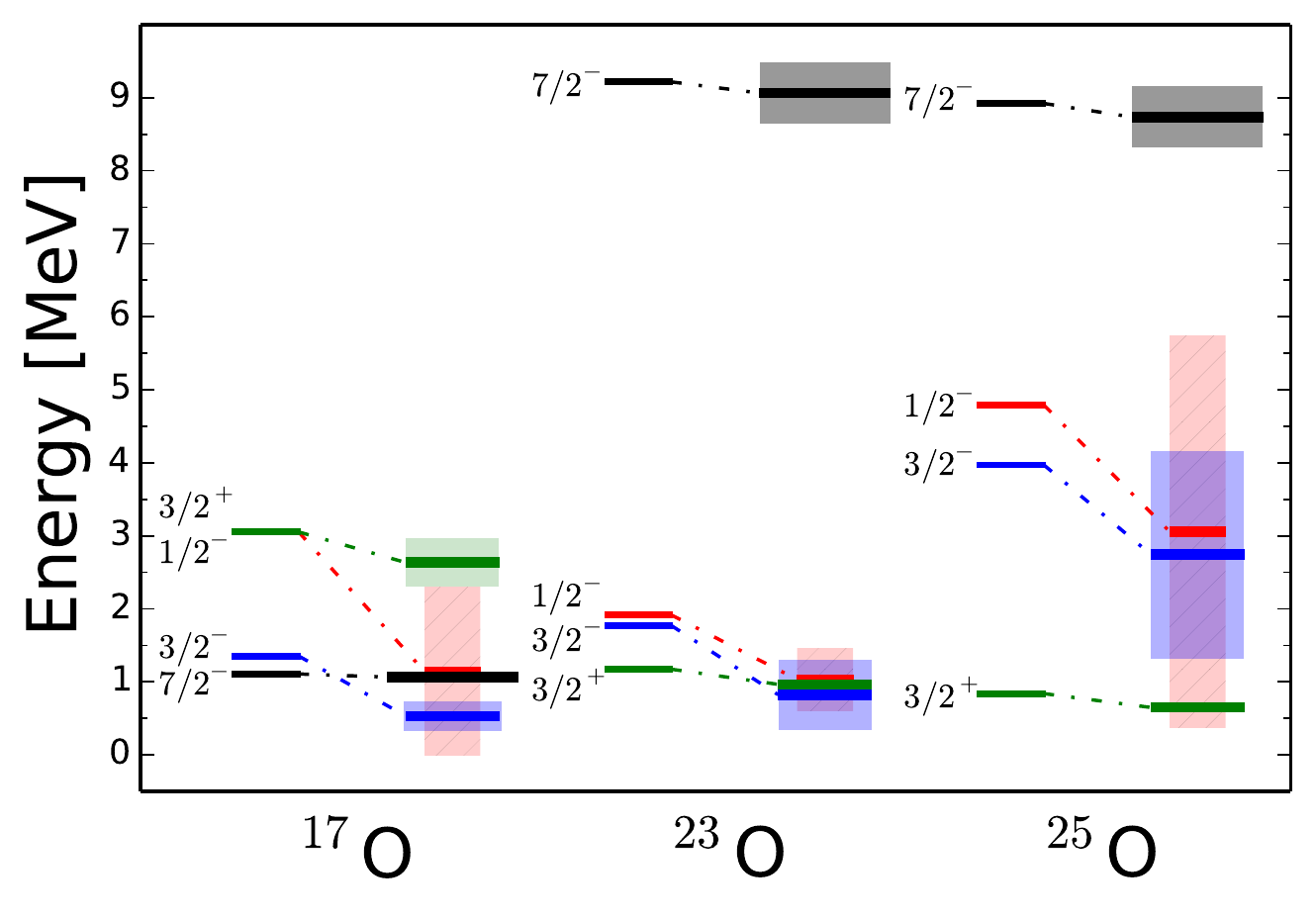}
\caption{(Color online) Energies (with respect to the
  particle-emission threshold) of unbound $J^\pi = {3/2}^+$,
  ${1/2}^-$, ${3/2}^-$ and ${7/2}^-$ states in the odd oxgyen isotopes
  $^{17,23,25}$O. For each nucleus the left group of states are
  computed using a harmonic-oscillator Hartree-Fock basis, while the
  corresponding right group states using a Gamow-Hartree-Fock
  basis. The role of continuum coupling on the individual states is
  indicated by the connecting dashed-dotted lines between the left and
  right group of states. The vertical bar on the right group of states
  gives the corresponding resonance width $\Gamma = -2\mathrm{Im}[E]$
  of each resonant state.}
  \label{Ox17_Ox25_res}
\end{figure*}

Figure~\ref{Ox17_Ox25_res} shows the energies of the unbound $J^\pi =
{3/2}^+,{1/2}^-,{3/2}^-,{7/2}^-$ states in the odd oxygen isotopes
$^{17,23,25}$O using a harmonic-oscillator basis and a Berggren basis
with respect to the particle-emission threshold. Here we only consider
the chiral interaction N$^3$LO + L(400), which is known to reproduce
spectra in the $sd$-shell. It is seen that the effect of the
continuum coupling on the $J^\pi = {3/2}^+$ and $J^\pi
= {7/2}^-$ resonant states is rather small, which is due to the large centrifugal
barrier ($l=2,3$) for these states and that these states are dominated
by one-particle radial wavefunctions of zero nodes, limiting thereby
their radial extension and making them quasi-bound within the
centrifugal barrier. On the other hand, the continuum coupling plays a
significant role on the $J^\pi =
{1/2}^-,{3/2}^-$ states.  These states are dominated by one-particle states
with a smaller centrifugal barrier ($l=1$) and radial wave functions
with one node that increases their radial extension. This results in a
larger sensitivity to the continuum, similar to what is found for
loosely-bound halo-states. In particular, we observe that the continuum is
responsible for inverting the order between the $J^\pi =
{3/2}^-,{7/2}^-$ states in $^{17}$O, while in $^{25}$O the $J^\pi =
{1/2}^-,{3/2}^-$ states are lowered by about 2~MeV and appear as very
broad resonances. These findings contrast the results obtained using
the harmonic-oscillator basis where no widths can appear. Such a large
width and a near degeneracy of the decay energies of the$J^\pi
= {1/2}^-,{3/2}^-$  states in $^{25}$O would present a significant challenge
for experiments targeting these states. We note that there are many
negative parity states in $^{17}$O, and for comparison to experiment
one would also need to know transition strengths. In $^{23,25}$O the
situation is simpler, and our results for the $J^\pi = {3/2}^+$
unbound states are in good agreement with the experimental data of
1.27~MeV \cite{elekes2007} and 0.77~MeV \cite{hoffman2008},
respectively. In these nuclei no data is available for negative parity
states.

Another interesting feature is that with increasing
neutron-to-proton ratio a gap between the $J^\pi = {3/2}^+$ state and
the $J^\pi = {1/2}^-,{3/2}^-$ states seems to be developing. One can
therefore speculate whether there is an 
$N=20$ (sub-) shell-closure in
$^{28}$O. However, this must be investigated in more detail by
addressing systematic uncertainties of the Hamiltonian and the
coupled-cluster method, as well as computing other observables such as
separation energies, pairing gaps, and the $2^+$ excited state, all being  
indicators of shell-closures.

\subsection{Unbound states in neutron-rich calcium isotopes} Recent
interest in understanding how shell structures evolve in calcium
isotopes has revealed several interesting and new features, such as
the newly suggested magic nuclei $^{52}$Ca and $^{54}$Ca with $N=32$
shell-closure \cite{wienholtz2013}, and $N=34$ (sub-) shell-closure
\cite{steppenbeck2013,steppenbeck2013b}. While 3NFs have been shown to
be important to explain these new (sub-) shell-closures
\cite{holt2012,hagen2012b}, coupled-cluster calculations of the odd
and neutron-rich isotopes $^{53,55,61}$Ca ~\cite{hagen2012b} based on
chiral nucleon-nucleon and schematic density-dependent 3NFs,
demonstrated that the role of the coupling to the particle continuum
significantly impacts and rearranges the order of states close to the
particle-emission thresholds. In particular, for the exotic nucleus
$^{61}$Ca the ground-state was computed to be a $J^\pi={1/2}^+$ state
dominated by configurations in the $s$ partial-wave, and with a
compressed and inverted order for the $J^\pi = {5/2}^+, {9/2}^+$
excited states in contrast to what one could expect from the standard
shell-model. In Ref.~\cite{hagen2013} coupled-cluster computations of
elastic neutron scattering on $^{60}$Ca were performed, and the
extracted scattering length for the $s$-wave phase-shifts was found to
be tens of fermi. The very large scattering length was used as input
to halo EFT \cite{bertulani2002,bedaque2003} in order to explore the
possible onset of Efimov physics in $^{62}$Ca.

The ``bunching'' of the $gds$ shell-model states in neutron-rich
calcium isotopes has also been observed in nuclear density-functional
and relativistic mean-field calculations, suggesting an onset of
deformation around $^{60}$Ca and with the dripline possibly extending
all the way out to $^{70}$Ca
\cite{nazarewicz1996,fayans2000,meng2002,erler2012}. In addition to
effects of the coupling to the continuum, deformation has also been
shown to impact and lead to a near degeneracy of shell-model states in
neutron-rich nuclei \cite{dobaczewski1994,hamamoto2012}. Shell-model
calculations based on phenomenological interactions also predicts a
bunching and inversion of the $d_{5/2}, g_{9/2}$ shell-model orbitals
in $^{61}$Ca \cite{lenzi2010}. Future experimental efforts at rare
isotope facilities promise to probe shell structure beyond
$^{54}$Ca. The heaviest calcium isotope to have been produced so far
is $^{58}$Ca \cite{tarasov2009} and data on $^{62}$Ti is currently
being analyzed at RIKEN. Confronting theory with data for the exotic
nucleus $^{60}$Ca and beyond might therefore not be too far in the
future.

In the remainder of this section we focus on the evolution of the
$J^\pi = {3/2}^+,{5/2}^+,{9/2}^+$ states in the odd calcium isotopes
$^{53,55,61}$Ca. In these exotic isotopes there are no  data for
these states. We again use the PA-EOM-CC method to compute states that
are dominated by one-particle excitations. Here we use the chiral
nucleon-nucleon interaction at N$^3$LO \cite{entem2003}, SRG evolved
to the cutoff $\lambda =1.8~\mathrm{fm}^{-1}$. The low-energy
constants of the leading-order chiral 3NF were adjusted to the binding
energies and radii of $^3$H and $^4$He, and within systematic
uncertainties reproduce the saturation point in nuclear matter
\cite{hebeler2011}. This chiral interaction was recently applied to
the computation of the binding energy, dipole polarizability, and the
neutron/proton root-mean-square point radii of $^{48}$Ca
\cite{hagen2015}. For this interaction, $^{48}$Ca was found to be
slightly overbound by about 4~MeV while the charge radius was
underestimated by about 0.2~fm. To construct a Gamow-Hartree-Fock
basis, we use the same model space as in Ref.~\cite{hagen2015}
(i.e. $N_{\rm max} = 14, E_{3\rm{max}}=16$ and $\hbar\Omega =
22$~MeV), and a Berggren basis for the $d_{3/2}, d_{5/2}, g_{9/2}$
partial waves.

\begin{figure*}[t]
\includegraphics[width=1.0\textwidth]{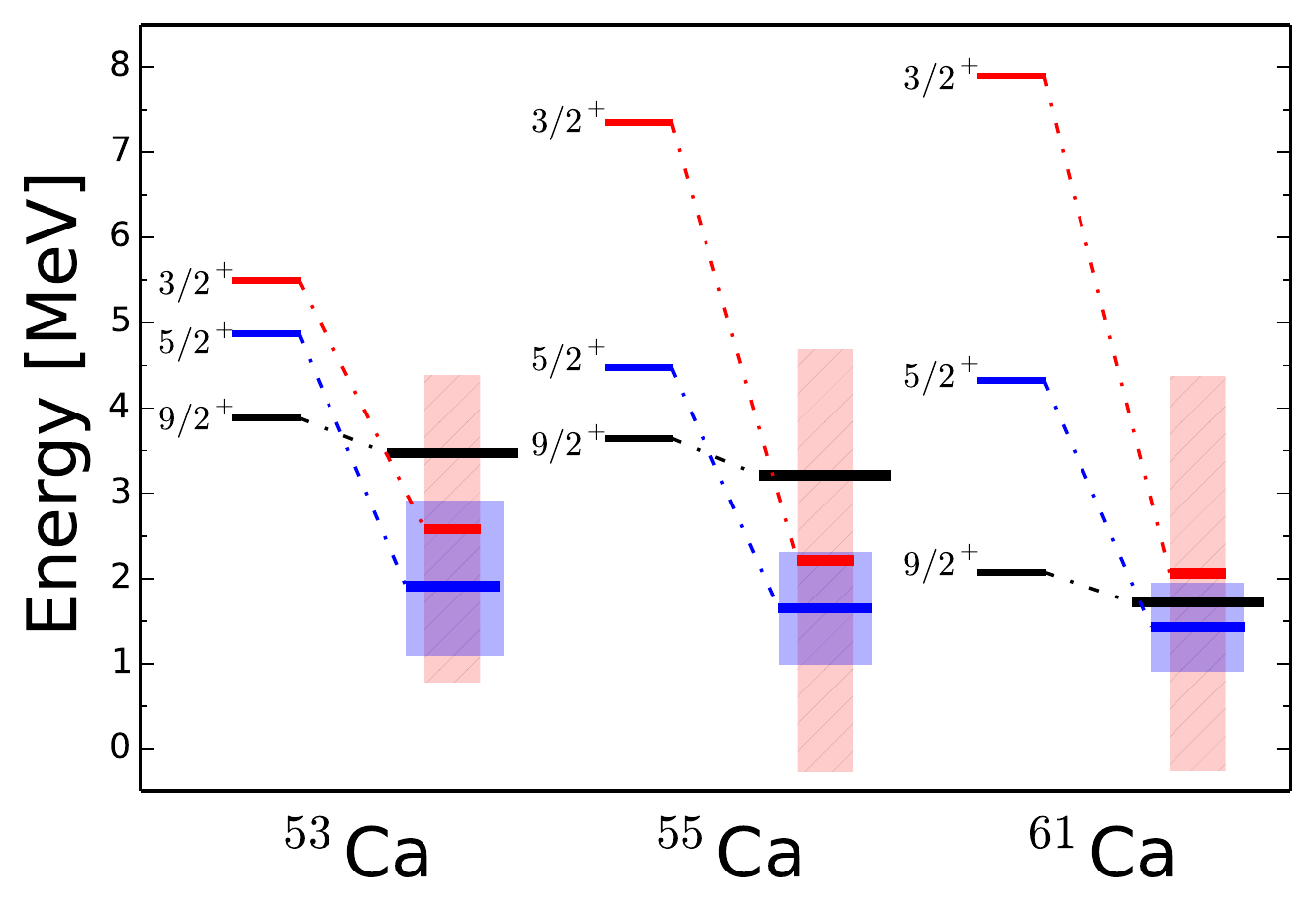}
\caption{(Color online) Energies (with respect to the
  particle-emission threshold) of unbound $J^\pi = {3/2}^+$,
  ${5/2}^+$ and ${9/2}^+$ states in the odd calcium isotopes
  $^{53,55,61}$Ca. For each nucleus the left group of states are
  computed using a harmonic-oscillator Hartree-Fock basis, while the
  corresponding right group states using a Gamow-Hartree-Fock
  basis. The role of continuum coupling on the individual states is
  indicated by the connecting dashed-dotted lines between the left and
  right group of states. The vertical bar on the right group of states
  gives the corresponding resonance width $\Gamma = -2\mathrm{Im}[E]$
  of each resonant state.}
  \label{Ca55_Ca61_res}
\end{figure*}

Figure~\ref{Ca55_Ca61_res} shows the $J^\pi = {3/2}^+,{5/2}^+,{9/2}^+$
unbound states in the odd calcium isotopes $^{53,55,61}$Ca, computed
in a Hartree-Fock basis built from harmonic oscillators only, and in a
Gamow-Hartree-Fock basis. We observe a striking feature, the coupling
to the continuum inverts the order of the $J^\pi =
{3/2}^+,{5/2}^+,{9/2}^+$ states, with the $J^\pi ={3/2}^+$ state being
a very broad resonance and with a decay energy located slightly above
the decay energy of the $J^\pi ={5/2}^+$ resonant state, which has a
considerable smaller width. Furthermore, in $^{61}$Ca we observe a
near degeneracy for the decay energies of the $J^\pi
={3/2}^+,{5/2}^+,{9/2}^+$ states. We recall that in \cite{hagen2013}
the ground-state of $^{61}$Ca was found to be a barely bound $J^\pi =
{1/2}^+$ state, dominated by $s$-waves. That the $J^\pi
={3/2}^+,{5/2}^+,{9/2}^+$ states are found to be close to the
threshold in $^{61}$Ca indicates that deformation and pairing will
play a significant role in the neutron rich calcium isotopes beyond
$^{61}$Ca. A proper theoretical description of isotopes beyond
$^{61}$Ca should therefore account for all these effects, that is,
continuum coupling, deformation and pairing, and many-body
correlations.

In the naive shell-model picture one could expect that the $J^\pi
={9/2}^+$ state, which is dominated by $g_{9/2}$ single-particle
configurations, should be lower in energy than the $J^\pi ={3/2}^+,
{5/2}^+$ states, which are dominated by $d_{3/2}, d_{5/2}$
single-particle configurations. As Fig.~\ref{Ca55_Ca61_res} shows,
this naive shell-model order is in fact realized if one neglects the
coupling to the particle-continuum and uses a harmonic-oscillator
basis only. This result is consistent with what was found in
Refs.~\cite{hagen2012b,hagen2013}, where a schematic 3NF was used. We
can understand the impact of the continuum on these unbound states by
employing a simple mean-field picture. In this picture the $J^\pi
={9/2}^+$ state is a pure $g_{9/2}$ single-particle state with a large
centrifugal barrier ($l=4$). Furthermore, its wave-function has no
radial nodes, localizing thereby the state within the barrier and making it
quasi-bound. Indeed, as one would expect, Fig.~\ref{Ca55_Ca61_res}
shows that the continuum has only a small impact on this state,
lowering it by a few hundred keVs compared to the harmonic-oscillator
basis, and with a width less than 10~keV. On the other hand, the
continuum has a significant impact on the unbound $J^\pi
={3/2}^+, {5/2}^+$ states. In the mean-field picture these states are pure
$d_{3/2}$ and $d_{5/2}$ single-particle states with a smaller
centrifugal barrier ($l=2$) and with wave-functions with one radial
node that pushes these states outside the barrier. The impact of the
continuum on these states is therefore significant, and similar to
what was found for the $J^\pi = {1/2}^-,{3/2}^-$ states in the oxygen
isotopes $^{17,23,25}$O.

One may speculate that the trend found for unbound states in
neutron-rich isotopes of oxygen and calcium may also be found in
neutron-rich nickel isotopes. Here one may expect a level inversion of
the unbound $J^\pi = {11/2}^-, {7/2}^-, {3/2}^-$ negative parity
states associated with the $0h_{11/2}, 1f_{7/2}, 2p_{3/2}$ mean-field
orbitals. Future experiments targeting spectroscopy in the exotic
nuclei $^{61}$Ca and $^{79}$Ni is therefore needed in order to fully
understand how shell-structure evolve at large neutron-to-proton
ratios and above particle decay thresholds.

\section{Emergence of nuclear deformation} 
Nuclear deformation is due to the emergent breaking of rotational
symmetry, that is the precursors of spontaneous breaking of rotational
symmetry in a finite
system~\cite{bohr1975nobel,yannouleas2007,papenbrock2014}. The key
signatures are rotational bands with energies $E\approx B I(I+1)$ for
states with angular momenta $I=0,2,4\ldots$, and strong collective
$B(E2)$ transitions between these states. Here, the rotational
constant $B$ is typically much smaller than the energy scales
associated with single-particle degrees of freedom~\cite{bohr1975}.
The collective nature and the low energy of the emergent rotational
constant $B$ make {\it ab initio} computations of such states
challenging. For the nucleus like $^{12}$C, the energies of Yrast
states were reproduced in the {\it ab initio}
computations detailed in Refs.~\cite{navratil2000,epelbaum2012,forssen2013,maris2014},
with accurate quadrupole moments in Ref.~\cite{epelbaum2012}.
Collective properties of other $p$-shell nuclei from {\it ab initio}
methods are addressed in
Refs.~\cite{dytrych2013,caprio2013,caprio2015}. Very recently, {\it ab
  initio} computations of Yrast states in $sd$-shell nuclei have
become possible~\cite{jansen2015,stroberg2015,hergert2015}. These
works are based on the recently developed effective shell-model
interactions from {\it ab initio} methods like coupled-cluster theory
and IMSRG \cite{bogner2014,jansen2014}.

The idea behind the coupled-cluster effective-interaction (CCEI)
method is to compute a non-perturbative shell-model interaction using
the coupled-cluster method and starting from state-of-the-art chiral
nucleon-nucleon interactions and 3NFs. In this approach, the
coupled-cluster method is used to compute spectra in nuclei with one,
and two nucleons outside a closed
core~\cite{gour2006,jansen2011,jansen2012,shen2014}. To be specific,
the computation of neutron-rich isotopes of carbon and oxygen can be
based on an effective neutron-interaction in the neutron $sd$-shell
model space.  Following Refs.~\cite{lisetskiy2008,dikmen2015}, {\it ab
  initio} computations of $^{14,15,16}$C and $^{16,17,18}$O are input
to the Okubo-Lee-Suzuki projection
technique~\cite{okubo1954,suzuki1980,suzuki1994,suzuki2000,kvaal2008},
yielding an effective interaction in the neutron $sd$ model space. The
symmetric orthogonalization~\cite{scholtz1992,navratil1996} yields an
effective interaction that is Hermitian. For each mass number one
obtains a new interaction because the intrinsic kinetic energy of the
underlying chiral Hamiltonian uses the mass number $A$ of the target
nucleus. Without any further tuning of parameters, the CCEI method
yields effective neutron-neutron interactions for oxygen and carbon isotopes with $^{16}$O and $^{14}$C as 
cores~\cite{jansen2014}, respectively. 
The resulting spectra are in good agreement
with experiment and comparable to the agreements obtained with
phenomenological Hamiltonians such as the USD \cite{brown1988} and the
WBP \cite{warburton1992} effective shell-model Hamiltonians.

\begin{figure*}[t]
\includegraphics[width=1.0\textwidth]{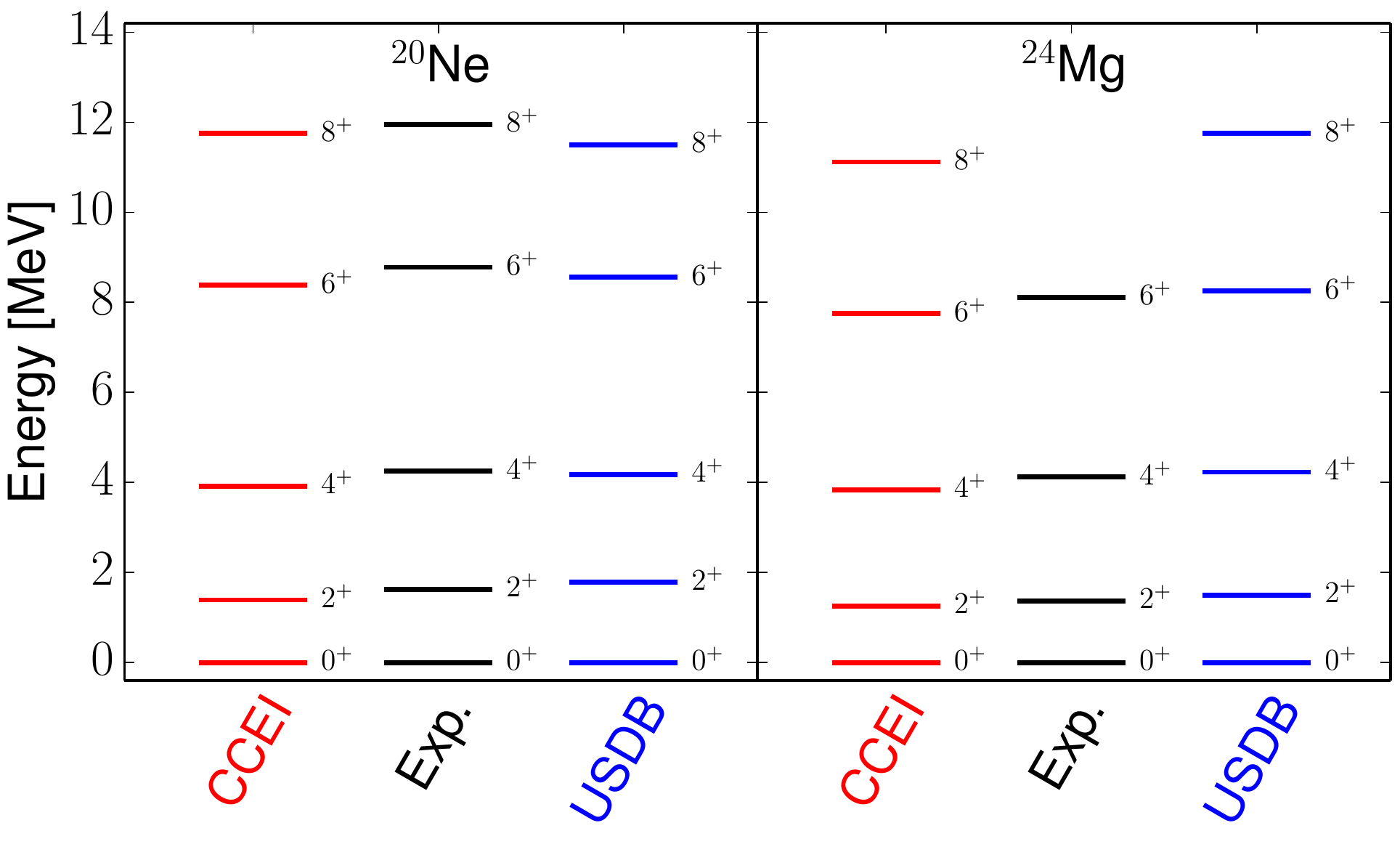}
\caption{(Color online) Yrast states in ${}^{20}$Ne and ${}^{24}$Mg
using CCEI and USDB compared to experimental values}
  \label{fig:ne20mg24yrast}
\end{figure*}

Recently, the CCEI has been extended to the full effective interaction
between neutrons and protons in the $sd$-shell valence space, taking
$^{16}$O as a core and performing {\it ab initio} coupled-cluster
calculations of $^{16,17,18}$O, $^{17,18}$F, and $^{18}$Ne.  These
calculations are based on the same chiral interaction used in
Ref.~\cite{jansen2014}, and the resulting effective shell-model
interaction was again truncated to one-body and two-body terms. In
contrast to Ref.~\cite{jansen2014}, this effective shell-model
interaction was constructed from a chiral interaction that uses the
mass number $A$ of the $^{16,17,18}$O, $^{17,18}$F, and $^{18}$Ne
nuclei. One obtains thereby an effective interaction for all
$sd$-shell nuclei that is independent of the nucleon number
$A$. Diagonalization of the resulting $sd$ shell Hamiltonian matrices
allowed the authors of Ref.~\cite{jansen2015} to compute rotational
bands of neon and magnesium isotopes.

Figure~\ref{fig:ne20mg24yrast} shows the ground-state rotational bands
in ${}^{20}$Ne and ${}^{24}$Mg and compares them to experimental data
and to results from the USDB interaction \cite{brown2006}. The
agreement of these yrast states to phenomenological interactions and
data is good.
 
The computation of rotational bands (and the accurate reproduction of
moments of inertia) are encouraging first results for {\it ab initio}
approaches to nuclear deformation. Our next steps will include the
computation of static and transition electric quadrupole moments as
well. The latter requires the development of effective (quadrupole)
operators that are calculated within the same theoretical framework.
Steps in these directions are under way and we expect that such
approaches will result in new insights into the theory of effective
charges.

\section{Summary and outlook: simplicity from complexity}
The physics of nuclei, with intricate forces yet to be completely
determined, two different fermionic species, many energy scales and
the absence of an external potential, generates a range and diversity
of behaviors that pose great challenges to a microscopic and
predictive description. Computing and understanding the wide range of
nuclear phenomena that emerge at different levels of complexity and
energy scales from the underlying forces and laws of motion is still
an open and unsettled issue. In the past decade, significant strides
have been made toward this goal. The description of heavier nuclei
using {\em ab initio} approaches has now become possible due to recent
development and advances in nuclear interactions,
renormalization-group techniques, many-body methods and an increasing
computing power that continues to scale according to Moore's law.

Several many-body methods are now available that exhibit a polynomial
scaling with mass number and are improvable in a systematic
way~\cite{dickhoff2004,hagen2010b,tsukiyama2011,soma2013,binder2013b,hergert2013b}.
This soft scaling combined with the exponential growth in computing
power allows us now to address heavier nuclei from first principles
for the first time. Modern nuclear forces, based on EFT concepts and
ideas from renormalization
group~\cite{epelbaum2009,bogner2010,machleidt2011}, and optimized to
reproduce data on selected light nuclei yielded accurate results and
even reliable predictions for a wide variety of
nuclei~\cite{otsuka2010,epelbaum2011,hagen2012b,wienholtz2013,hergert2013}. An
EFT interaction that was constrained to radii and binding energies of
isotopes up to oxygen~\cite{ekstrom2015} accurately
reproduced the charge densities in $^{40,48}$Ca and yielded a prediction
of the neutron distribution of $^{48}$Ca \cite{hagen2015}.  The
methodological advances in constructing single-particle and
many-particle basis functions for continuum properties, allow us to
study nuclei close to the limits of stability.  Furthermore, as we
illustrated here, we are now in a position where we can compute
effective Hamiltonians (including single-particle energies and
two-body interactions) for a single major shells. These effective
interactions are without empirical adjustments and led to the
computation of spectra of $sd$-shell nuclei. The onset of deformations
in neon and magnesium isotopes were studied recently
\cite{jansen2015,stroberg2015} and reviewed in the previous
section. These results demonstrate the potential and capabilities of
present many-body methods.

Simple patterns in nuclei such as nuclear saturation, rotational and
vibrational spectra, Borromean properties and halo-like density
distributions properties, and shell-structure are now being approached 
from microscopic many-body calculations. We are reaching a point
where few-body and mean-field models specifically aimed at describing
such emergent phenomena can be complemented by more microscopic and
refined models. This is not to say that these simpler models will be
replaced by ever more complex and complicated quantum-mechanical
many-body descriptions. They will continue to play an important role
in the discovery process of new phenomena, and in their interpretation
and understanding in terms of more intuitive pictures and ideas.

The microscopic description of nuclei from first principles highlights
various aspects of the nuclear many-body problem, ranging from the
optimization of low-energy coupling constants of the nuclear
interactions, the role of 3NFs, the role of high-momentum modes and
many-body correlations, and to the treatment of the atomic nucleus as
an open-quantum system with a strong coupling to the particle
continuum and decay channels. Depending on the energy scales that are
relevant for the description of the specific nuclear property and its
origin, these components vary in their importance. For example,
including the coupling to the particle continuum is not necessary for
strongly bound nuclei, but it plays a crucial role on evolution of
shell-structure in neutron-rich oxygen and calcium isotopes.  Another
example is the important role of 3NFs on nuclear saturation and on
shell closures, while its role on \emph{ab initio} calculations of
rotational bands in the deformed nuclei $^{20}$Ne and $^{24}$Mg seems
less prominent. Simple models of the nucleus may therefore often guide
us in deciphering the importance of these different components.

We end this paper by pointing out several open challenges and exciting
avenues.  The recent work \cite{binder2015b} with interactions from
EFT tailored to a harmonic oscillator basis, might hold some promise
for performing accurate {\it ab initio} calculations for heavier
nuclei and address issues related to nuclear shell-structure and
saturation over the entire nuclear chart. The technical capabilities
are now in place~\cite{binder2013b}, and the onus is on the
optimization of accurate interactions. Further development of
open-shell methods such as CCEI and valence-space IM-SRG might open up
a path for {\it ab initio} studies of neutrinoless double-beta decay
in nuclei such as $^{76}$Ge and $^{136}$Xe. Along similar lines, a
derivation of effective operators and interactions for heavier nuclei
opens up possibilities for studies of  electroweak
interactions in a nuclear medium.

These studies and the pertinent formalism can also be extended to
infinite matter, of great interest for studies of dense objects like
neutron stars. Reliable many-body derivations of the equation-of-state
and effective interactions and operators for dense nuclear matter, are
of great importance for far ranging issues like neutrino oscillations
and the synthesis of the elements. Another topic which is of crucial
importance for coming experimental programs is the link between the
structure and the dynamics of nuclei, with topics like reactions in
medium-mass nuclei and the derivation of for example optical
potentials using present many-body methods.

Finally we would like to point out an unresolved challenge to our
many-body methods. For any \emph{ab initio} calculation to be truly
meaningful, uncertainties at all levels of the calculation needs to be
quantified. There are recent advancements in the quantification of
systematic and statistical uncertainties of the employed Hamiltonian
\cite{carlsson2015,epelbaum2015,perez2015} and of employing Bayesian
statistics in nuclear
EFTs~\cite{schindler2009,furnstahl2014c,furnstahl2015,coelloperez2015b}. Propagation
of these uncertainties from the underlying Hamiltonian to the nuclear
many-body problem are underway \cite{ekstrom2015b,lynn2015}. Besides
uncertainties in the employed Hamiltonian, many-body methods auch as
coupled-cluster theory rely on an expansion using a finite
single-particle basis. Finite basis truncations introduces another
uncertainty which needs to be quantified. Recently formulas for
infrared and ultraviolet extrapolations of binding energies, radii and
quadrupole moments based on a harmonic oscillator basis in few- and
many-body systems were derived
\cite{furnstahl2012,coon2012,more2013,furnstahl2014,furnstahl2014b,konig2014,wendt2015,odell2015}.
We also mention that for specific systems like quantum dots using
effective interactions, precise error estimates for a given truncation
of the harmonic oscillator basis may also be given, see for example
Ref.~\cite{kvaal2009}. Finally, there is a systematic uncertainty
related to the specific truncation of the applied many-body method.
This uncertainty is perhaps the most challenging and difficult to
address, with first steps made for the coupled-cluster
truncation~\cite{kutzelnigg1991,rohwedder2013}. Although most many-body methods are
subject to systematic improvements and controlled approximations, a
rigorous theory for the convergence with respect to truncation level
is an unresolved and open problem.

\ack We thank W. Nazarewicz for providing us with
Fig.~\ref{NNLOsat_strategy}. This work was supported by the Office of
Nuclear Physics, U.S. Department of Energy (Oak Ridge National
Laboratory), DE-SC0008499 (NUCLEI SciDAC collaboration), the Field
Work Proposal ERKBP57 at Oak Ridge National Laboratory, the grant
No. DE-FG02-96ER40963 (University of Tennessee), the National Science
Foundation Grant No.~PHY-1404159 (Michigan State University) and by
the Research Council of Norway under contract
ISP-Fysikk/216699. Computer time was provided by the Innovative and
Novel Computational Impact on Theory and Experiment (INCITE)
program. This research used resources of the Oak Ridge Leadership
Computing Facility, which is a DOE Office of Science User Facility
supported under Contract DE-AC05-00OR22725, and used computational
resources of the National Center for Computational Sciences and the
National Institute for Computational Sciences.

\section*{References}

\bibliographystyle{iopart-num}

\providecommand{\newblock}{}

\end{document}